\documentclass[a4paper,fleqn,usenatbib]{mnras}

\usepackage[T1]{fontenc}
\usepackage{ae,aecompl}

\usepackage{graphicx}
\usepackage{amsmath}
\usepackage{amssymb}
\usepackage{float}
\usepackage{enumerate}  
\usepackage{lipsum}
\usepackage{gensymb}
\usepackage{graphicx}
\usepackage{longtable}
\usepackage{rotate}
\usepackage{rotating}
\usepackage{mathtools}
\usepackage{pdflscape}
\usepackage{scalerel}
\usepackage{times}
\usepackage{verbatim}
\usepackage{tablefootnote}
\usepackage{subfig}
\usepackage{csquotes}
\newif\ifAMStwofonts

\def\nustar{{\it NuSTAR}}

\def\suzaku{{\it Suzaku}}

\def\athena{{\it Athena}}
\def\xrism{{\it XRISM}}

\def\xmm{{\it XMM-Newton}}

\def\mrk335{{Mrk~335}}

\def\feka{{Fe~K$\alpha$}}

\def\fexxv{{Fe~\textsc{xxv}}}
\def\fexxvi{{Fe~\textsc{xxvi}}}

\def\cm{{\rm\thinspace cm}}
\def\erg{{\rm\thinspace erg}}
\def\eV{{\rm\thinspace eV}}

\def\kev{{\rm\thinspace keV}}

\def\s{{\rm\thinspace s}}

\def\ps{{\rm\thinspace s^{-1}}}

\def\ergps{\hbox{$\erg\s^{-1}\,$}}

\def\pscm{\hbox{$\cm^{-2}\,$}}

\title[Iron line profiles of \suzaku\ Seyfert 1s]{A \suzaku\ sample of unabsorbed narrow-line and broad-line Seyfert 1 galaxies: II. Iron emission and absorption}

\author[S. G. H. Waddell \& L. C. Gallo]{
S. G. H. Waddell$^{1}$\thanks{E-mail: swaddell@mpe.mpg.de}
and L. C. Gallo$^{1}$
\\
$^{1}$Department of Astronomy \& Physics, Saint Mary's University, 923 Robie Street, Halifax, Nova Scotia, B3H 3C3, Canada \\
}

\date{Accepted XXX. Received YYY; in original form ZZZ}

\pubyear{2021}

\begin{document}
\label{firstpage}
\pagerange{\pageref{firstpage}--\pageref{lastpage}}
\maketitle

\begin{abstract}
A sample of 22 narrow-line Seyfert 1 (NLS1) and 47 broad-line Seyfert 1 (BLS1) galaxies observed with \suzaku\ is used to examine the Fe K band properties of each group. Three different models are used to examine the presence of: narrow neutral \feka\ line at $6.4\kev$ and ionised \fexxv\ and \fexxvi\ emission lines (model A); a broad emission feature at around $6-7\kev$ (model B); and an absorption edge at $\sim7.1\kev$ (model C). In all three models, the neutral \feka\ line is weaker (lower luminosity and equivalent width) in NLS1s than in BLS1s. Model (B) also finds a more significant broad component (larger equivalent width) in NLS1s than in BLS1s. The feature does not appear to be an artefact of steeper spectra in NLS1s, but rather an intrinsic property of these sources. From model (C), the optical depth of the absorption edge appears comparable between the two samples. When comparing the absorption with the emission line properties, NLS1s seem to exhibit a lower ratio of emission-to-absorption of iron than BLS1s, and than expected based on the fluorescence yield. The observed differences may arise from different torus geometries (e.g. larger opening angle in NLS1s), and/or additional sources of Fe K emission and absorption in NLS1s beyond pure fluorescence (e.g. originating in the disc and broad line region).
\end{abstract}




\section{Introduction}
\label{sect:intro}
Examining the X-ray emission from Active Galactic Nuclei (AGN) allows for the study of some of the most extreme processes occurring at the innermost regions of the system. In particular, the study of type-1 AGN, or Seyfert 1s, allows a direct view of the central engine (\citealt{Antonucci+1993}). The spectra of these objects is characterised primarly by a power law component, believed to originate from the Comptonisation of UV seed photons from the disc by a hot corona located above the disc (e.g. \citealt{Haardt+1991,Haardt+1993}). This component typically dominates the spectrum between $2-10\kev$. Many AGN also feature a strong soft excess, an excess of photons below $\sim2\kev$; and a Compton hump peaking at $20-30\kev$ produced via Compton down-scattering of X-ray photons in an optically thick medium (e.g. the inner layers of the disc or the distant torus), or in an optically thin medium where the peak is shifted to lower energies (e.g. \citealt{MY+2009}).

A well studied region of the X-ray spectra of AGN is the iron band, around $4-8\kev$. Numerous emission and absorption features of varying width can be seen here, most prominently the neutral \feka\ emission line at $6.4\kev$ and accompanying absorption edge at $7.1\kev$. Ionised emission lines from Helium-like iron (\fexxv) and Hydrogen-like iron (\fexxvi) and their accompanying edges may also be present. Features may be broadened if they are produced in the accretion disc, or shifted if they are inflowing or outflowing. It is expected that all or some combination of these features are present in all AGN, and notably, it is likely that some combination of narrow lines and broadened emission is seen. The characterisation of this complex iron line region is very important to build an understanding of the presence of iron in the vicinity of the AGN, the illumination of iron (in the torus, broad-line region or accretion disc) by X-ray photons, and the geometry of the torus and inner region of the AGN.

There are many previous works which make use of the \suzaku\ satellite to study the iron line profiles of AGN. Individual sources have been studied to probe the shape of the spectrum and study the \feka\ region in detail, including NGC 3227 (\citealt{Markowitz+2009b}), NGC 4593 (\citealt{Markowitz+2009b}), and MCG -2-58-22 (\citealt{Rivers+2011}). Samples of sources have been examined, to study global properties of neutral and ionised narrow emission features (\citealt{Fukazawa+2011,Fukazawa+2016,Mantovani+2016}) and properties of a broad component interpreted as relativistic broadening of the iron line (\citealt{Reeves+2006,Patrick+2011,Patrick+2012}). Other works have also studied similar properties using \xmm, examining both individual spectra (e.g. \citealt{Nandra+2007,Zhou+2010}) and stacked spectra (e.g. \citealt{Liu+2015,Liu+2016}) to determine the average properties of samples.

In a previous work (\citealt{WG20}; hereafter WG20), a sample of 69 Seyfert 1 galaxies observed with \suzaku\ was presented (see tables 1 and 2 of WG20 for a complete list of objects in the sample and some relevant properties). Objects with redshifts $z<0.5$ and host galaxy column densities $<10^{22}$\pscm were selected to ensure adequate modelling of the soft excess. These objects were then divided into 22 narrow-line (NLS1) and 47 broad-line (BLS1) AGN, classified based on the FWHM of the optical H$\beta$ line (where NLS1 galaxies have FWHM less than $2000~\textrm{km}\ps$; see \citealt{Osterbrock+1985} and \citealt{Goodrich+1989}). The X-ray spectral properties of the sample were analysed, and several differences between the two classes were found. NLS1 galaxies showed steeper photon indices, stronger soft and hard excesses, and a strong correlation between the hard and soft excesses not found in BLS1s. 

All sample selection and data reduction techniques are described in sections 2 and 3 of WG20. Here, the same sample will be examined, with a focus on the iron emission in the $\sim4-8\kev$ band, and absorption properties in the $7-20\kev$ band. This work expands upon previous works by including a large sample of Seyfert 1 AGN, improved fitting techniques (including the use of optimal binning and C-statistics) and comparing a wide range of parameters between NLS1s and BLS1s. In Section~\ref{sect:lines}, three distinct models for the iron emission and absorption features are applied to each spectrum, and properties are compared. A discussion of results and comparison between the NLS1 and BLS1 parameters are given in Section~\ref{sect:disc}, and conclusions are drawn in Section~\ref{sect:conclusion}.

\section{Emission and absorption modelling}
\label{sect:lines}

\subsection{Modelling procedure}
\label{sect:model}

All spectral fits are performed using {\sc xspec v12.9.0n} (\citealt{Arnaud1996}) from {\sc heasoft v6.26}, and abundances are taken from \cite{Wilms+2000}. Spectra from the \suzaku\ XIS detectors have been grouped using optimal binning (\citealt{Kaastra2016}), using the {\sc ftgrouppha} tool, while data from the PIN have been grouped to a minimum of 20 counts per bin. To maximize spectral data for each source, the XIS0 and XIS3 spectra from all epochs are combined to form a single spectrum, which is then modelled. The XIS background spectrum has been modelled, rather than subtracted for each spectrum. This procedure is described in more detail in WG20. Spectral fitting is evaluated using C-statistic (\citealt{Cash1979}).

As the focus of this work is on the Fe~K band, and to minimize the effects of absorption, only the $3-10\kev$ band is considered for the XIS data. All spectral fits include the Galactic column density taken from \cite{Willingale+2013}, though this rarely has a significant effect on the spectral shape above $3\kev$.

\begin{figure}
	\centering
	\includegraphics[width=\columnwidth]{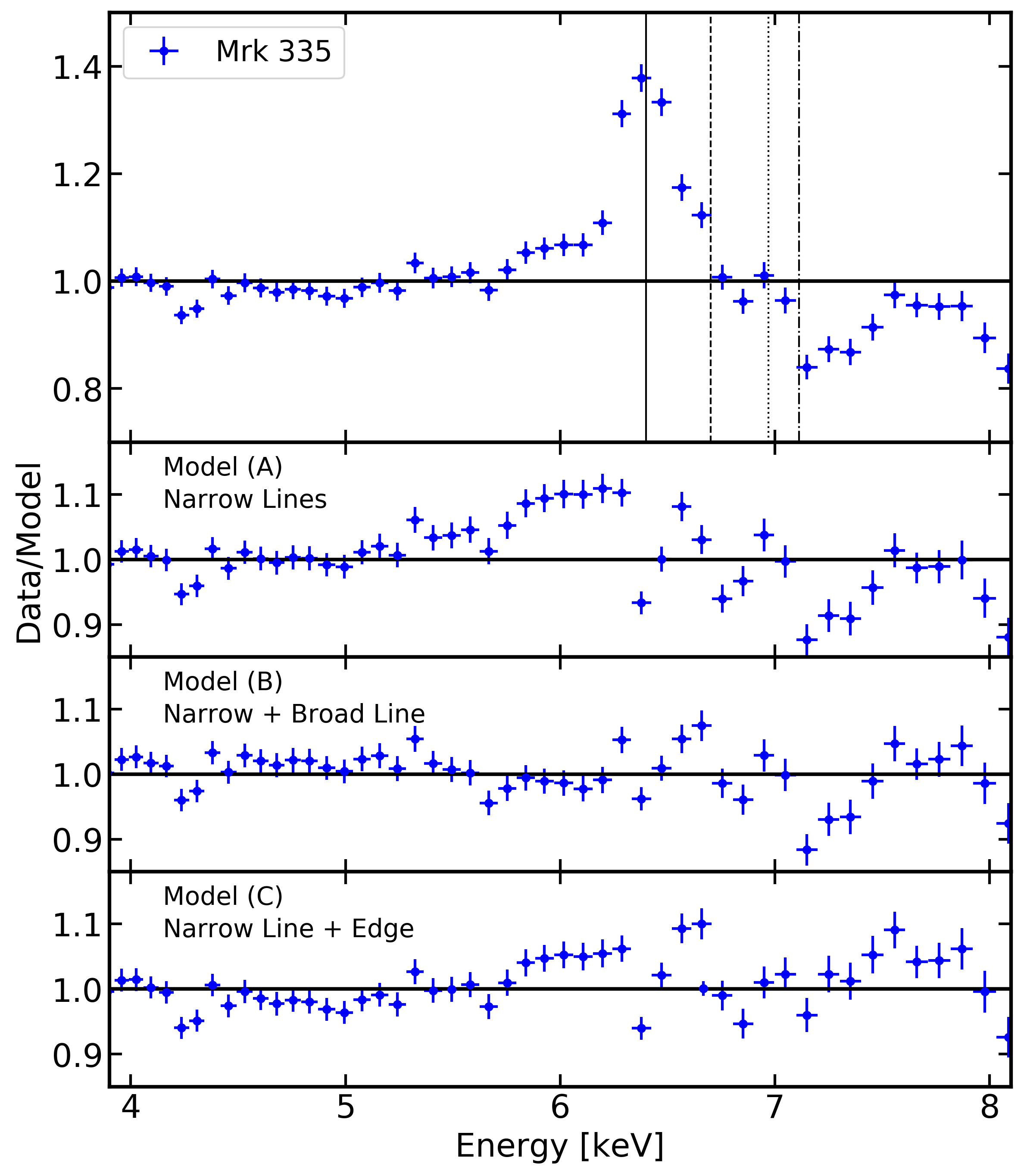}
	\caption{Iron emission profiles for Mrk 335, shown in the $4-8\kev$ band for clarity. Top panel: power law fit to the $3-10\kev$ band. For reference, the energies $6.4$, $6.7$, $6.97$ and $7.1\kev$ are shown as solid, dashed, dotted, and dash-dot lines, respectively, where we expect to see iron features. Second panel: Model (A), a power law plus three narrow lines corresponding to neutral \feka, \fexxv\ and \fexxvi\ at $6.4$, $6.7$ and $6.97\kev$, respectively. Third panel: Model (B), a power law, a narrow line at $6.4\kev$, and a broad line with energy and width left free to vary. Bottom panel: Model (C), a power law, narrow iron line, and absorption edge with energies free to vary. }
	\label{fig:m335fe}
\end{figure}

Three distinct iron emission models are considered to model the iron band. The residuals (data/model) of these models applied to the NLS1 galaxy Mrk 335 (e.g. \citealt{Gallo+2015,Gallo+2019m335}) are shown in Fig.~\ref{fig:m335fe}. The top panel shows the residuals to a power law fit, and clear emission and absorption features are present. Model (A), shown in the second panel, includes a power law plus three narrow Gaussian lines. In model (B) (third panel), the ionised iron lines are removed and a broad line is added, with the width and energy left free to vary.

Finally, in model (C) (bottom panel), the absorption edge is modelled. A narrow line to account for neutral \feka\ emission is also included. To accurately measure the photons absorbed by the edge, the PIN data for each source are included in the spectral fit, where available. As many sources exhibit strong signatures of a Compton hump, an additional broad Gaussian is added to the model, with the energy and width frozen to $20\kev$ and $5\kev$, respectively. This model was selected for consistency with WG20, where it was shown that the broad Gaussian accurately approximated the shape of the Compton hump. Additionally, it was shown that using a broad Gaussian rather than a more physical model allows for independent measurements of the narrow line, edge, and Compton hump to be made and ensures adequate characterisation of all features. 

Mrk 335 is typical of what is seen in most high quality spectra in the sample. In the top panel, the power law is capable of fitting the spectrum in the $4-5$ and $7-8\kev$ bands, but there are clear excess residuals in the $5-7\kev$ regime. In the second panel, the addition of a $6.4\kev$ line and two narrow ionised lines improves the fit, with $\Delta C = 797$ for three additional free parameters. An excess is still visible in the $5-6.4\kev$ band. In the third panel, the addition of a broad line significantly improves the residuals at lower energies than the \feka\ line, as is the case for many of the sources in the sample. When compared to the power law fit, a $\Delta C$ of $999$ for four additional free parameters is obtained. Finally, in the fourth panel, the edge models the dip seen at $\sim7.1\kev$ in the residuals of the above panels.

This modelling procedure is repeated for every NLS1 and BLS1 in our sample, and the results described in Sections~\ref{sect:nline}, \ref{sect:bline}, and \ref{sect:edge}. A summary of all models, free parameters, and the implementation in {\sc xspec} is given in Table~\ref{tab:models}. It is important to again note that all three processes are likely present in AGN spectra, but most objects do not require a combination of models for the given data quality. We note also that the iron line normalisations are fit using the cflux component as opposed to allowing the line normalisation. As a result, absorption features cannot be modelled using this method. This may introduce some bias, however, it restricts the physical interpretation to emission only, which is the main consideration of this work.

\begin{table*}
	\caption{Abbreviated model names as referenced in this work as well as their implementation in {\sc xspec}. $^{1}$ only free to vary if it improves the C-statistic.}
	\label{tab:models}
	\resizebox{\textwidth}{!}{%
		\begin{tabular}{lll}
			\hline
			Name & {\sc xspec} implementation & Free parameters\\
			\hline
			Baseline & TBabs(cflux*powerlaw) & $\Gamma$, PL flux \\
			Model (A) & TBabs(cflux*zgauss$_{6.97}$ + cflux*zgauss$_{6.7}$ + cflux*zgauss$_{6.4}$ + cflux*powerlaw) & $\Gamma$, PL flux, $6.4\kev$, $6.7\kev$ and $6.97\kev$ fluxes \\
			Model (B) & TBabs(cflux*zgauss$_{\rm broad}$ + cflux*zgauss$_{6.4}$ + cflux*powerlaw) & $\Gamma$, PL flux, $6.4\kev$ flux, broad line flux, energy, $\sigma$ \\
			Model (C) & constant*TBabs(gaussian$_{\rm CH}$ + cflux*zgauss + zedge*powerlaw) & $\Gamma$, PL flux, Compton hump norm, line flux, line \\
			& & energy$^{1}$, edge energy$^{1}$, $\tau$ \\
			\hline
			\hline
		\end{tabular}
	}
\end{table*}

\subsection{Model (A) - Ionised emission lines}
\label{sect:nline}

First model (A), a power law plus three Gaussian lines with $1\eV$ width at $6.4$, $6.7$ and $6.97\kev$ is applied to each spectrum. The narrow line widths imply that the origin is far from the black hole, likely in the broad-line region (BLR) or inner layers of the torus (\citealt{Matt+1996}; \citealt{Bianchi+2002}). The $6.7$ and $6.97\kev$ features arise from highly ionised iron, while the $6.4\kev$ line originates in a neutral medium. The average equivalent widths (EW) of the emission features are found to be $\sim0.02\kev$ for the ionised lines and $\sim0.05\kev$ for the $6.4\kev$ feature.

The addition of these features generally improves the fit when compared to a simple power law model, with $\Delta C$ of $50-250$ for three additional free parameters. This implies the presence of strong iron emission features in many of the objects in the sample, in agreement with numerous previous works (e.g. \citealt{Reeves+2006,Nandra+2007,Patrick+2012}). This fit improvement is seen in both the NLS1 and BLS1 samples, indicating that emission features are present in both samples as expected. 

\begin{figure*}
	\centering
	\subfloat[]{
		\includegraphics[width=57mm]{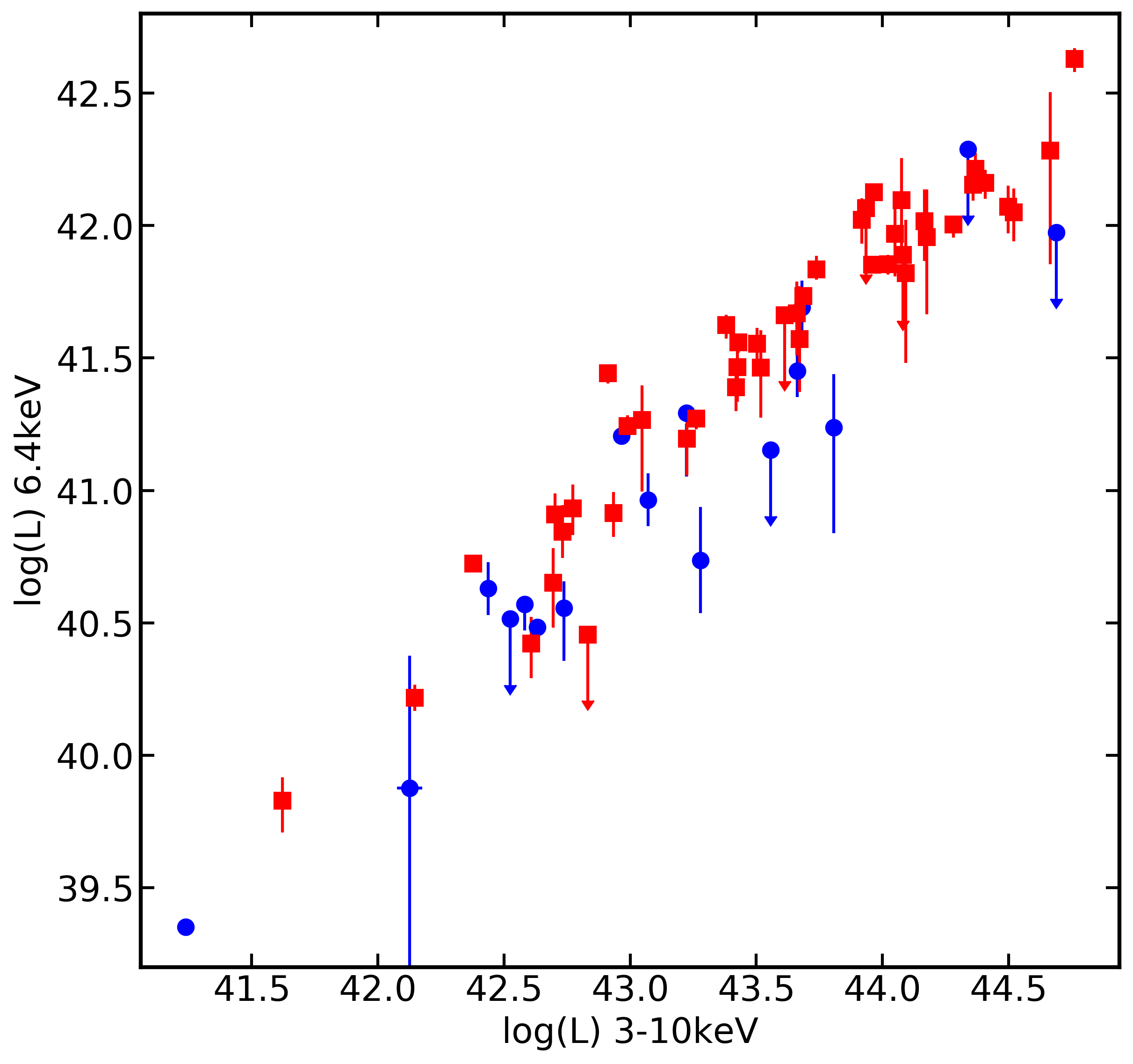}
	}
	\hspace{2mm}
	\subfloat[]{
		\includegraphics[width=55mm]{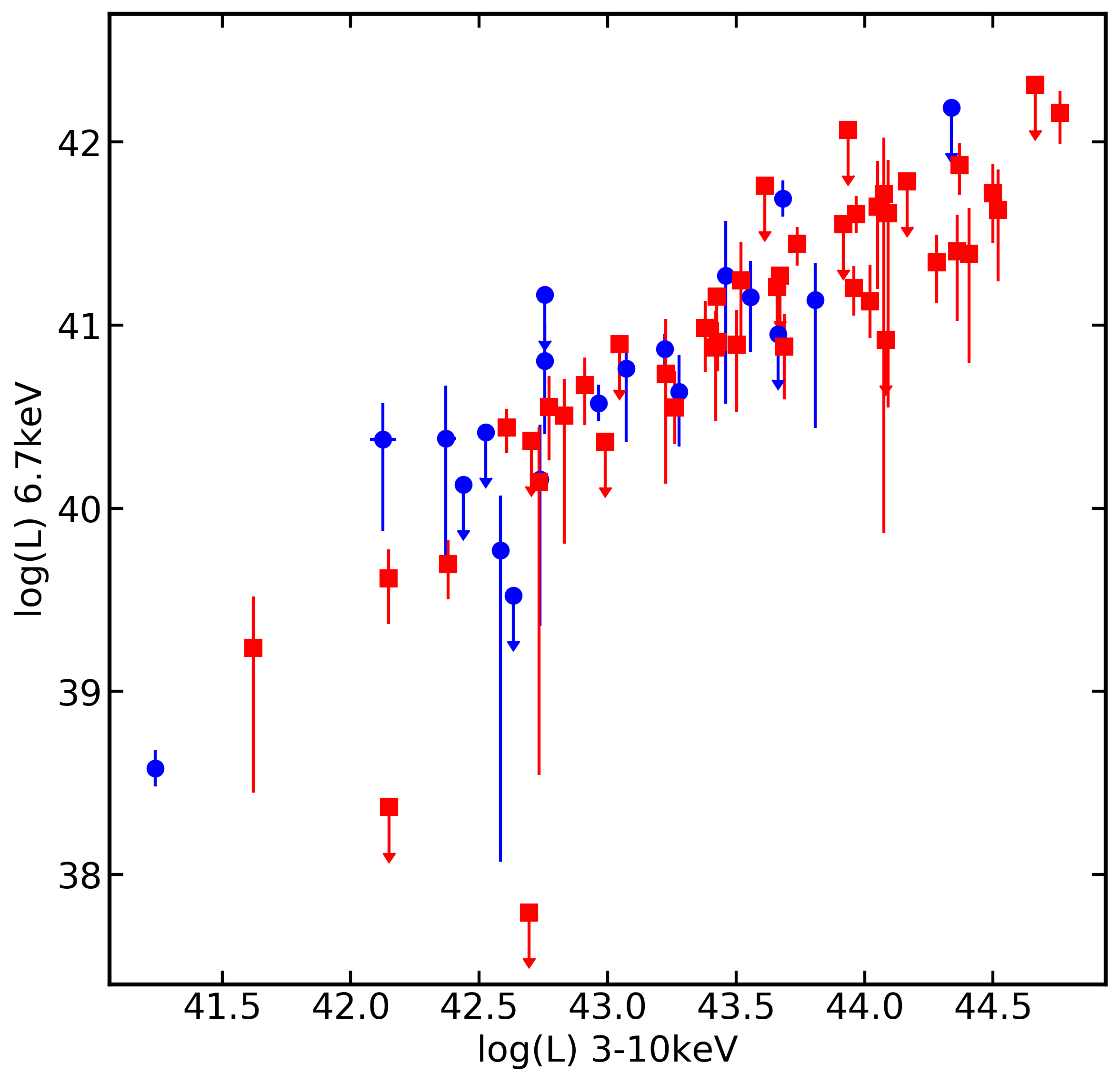}
	}
	\hspace{2mm}
	\subfloat[]{
		\includegraphics[width=55mm]{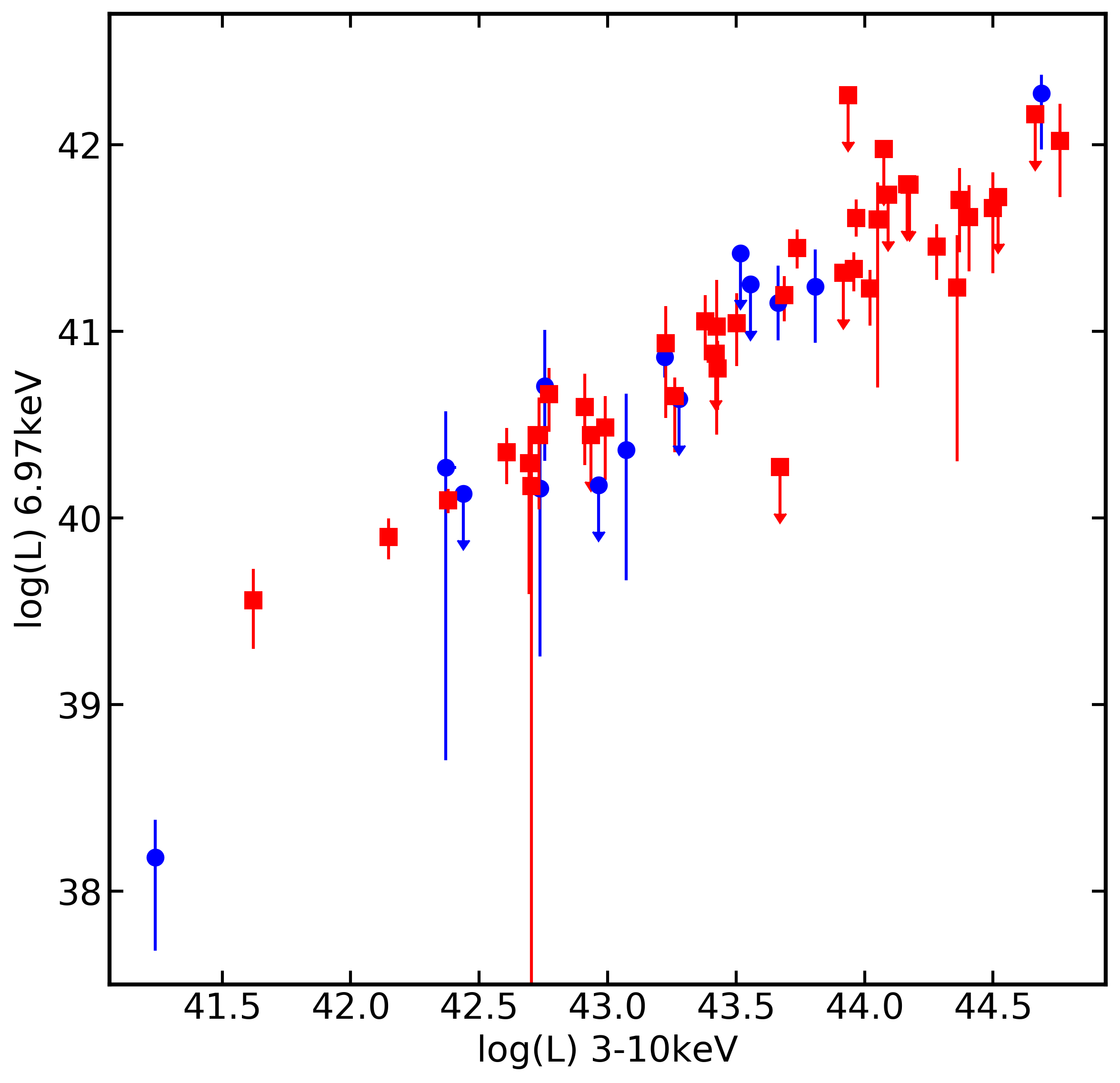}
	}
	\caption{Line luminosities (\ergps) with respect to the continuum luminosity for each source in the sample. Both values are shown as log values for clarity. NLS1 sources are shown as blue circles, and BLS1 sources are shown as red squares. Upper limits are indicated with arrows. Panel (a) shows the $6.4\kev$ line, panel (b) shows the $6.7\kev$ line and panel (c) shows the $6.97\kev$ line. }
	\label{fig:nlsct}
\end{figure*} 

To compare the two samples it is useful to consider the detection rate for each feature; that is, how many objects of each type have non-zero line luminosity at the 90 per cent confidence level. Examining first the narrow \fexxvi\ feature at $6.97\kev$, 27/47 (57\%) of the BLS1s and 9/22 (41\%) of the NLS1s show significant detections. Many NLS1s have lower signal to noise at high energies, which may explain the slightly lower percentage of NLS1s with significant \fexxvi\ features. The values for the $6.7\kev$ \fexxv\ lines are more similar, with 30/47 (64\%) of the BLS1s and 15/22 (68\%) of the NLS1s showing significant detections. These fractions are similar to those found by \cite{Patrick+2012} using a sample of 46 Seyfert 1 AGN observed with \suzaku. Furthermore, ionised iron emission features have been reported in both NLS1 and BLS1 AGN (e.g. \citealt{Patrick+2012,Mantovani+2016,Ehler+2018,Soph}). Overall, this suggests that the presence of ionised iron emission is similar for NLS1 and BLS1 galaxies. 

Such similarities do not appear when examining the neutral \feka\ line at $6.4\kev$. While 42/47 (89\%) of BLS1s show significant detections, this is true for only 12/22 (55\%) of NLS1s.  These fractions are similar to those observed in \cite{Zhou+2010}, who study a sample of \xmm\ spectra and find that 46\% of NLS1s and 82\% of BLS1s have significant detections of the narrow \feka\ line, and thus conclude that NLS1s have weaker narrow \feka\ lines. This may indicate that narrow \feka\ features are less common in NLS1s than in BLS1s, or that they are simply weaker and harder to detect, on average, in NLS1s.

Examining now the distributions of the line luminosities and equivalent widths, NLS1 galaxies have lower median luminosity and equivalent width than their BLS1 counterparts. To characterise this, a KS test is used, implemented in {\sc python} using {\sc scipy.stats.ks\_2samp()}. The KS test assesses the probability that two samples are drawn from the same parent distribution, and returns a p-value. In general, p-values less than 0.05 suggest a difference with $>99.5\%$ confidence, and these values should be considered significant, or at least worthy of interest. For the $6.4\kev$ line luminosity distributions, the KS test returns a p-value of 0.0057, implying that the distributions are different at the $\sim99.5$ per cent level. When considering the equivalent width of the line, a higher p-value of 0.094 is obtained. Little can be drawn from the difference in equivalent width, but considering the lower lines luminosities combined with the lower detection rate of narrow $6.4\kev$ lines in NLS1s, it can again be concluded that NLS1s have weaker neutral \feka\ lines. This is in agreement with the results of the broad-band modelling performed in WG20.

Such differences are not found when examining the ionised \fexxv\ and \fexxvi\ emission parameters The KS test finds p-values of 0.076 and 0.032 for the luminosity and equivalent width distributions of the $6.7\kev$ line, respectively; and  p-values of 0.21 and 0.42 for the luminosity and equivalent width distributions of the $6.97\kev$ line, respectively. This implies that the properties of \fexxv\ and \fexxvi\ lines are very similar between the two samples.

To further investigate potential differences observed for the neutral \feka\ line, Fig.~\ref{fig:nlsct} shows the distributions of $6.4\kev$, $6.7\kev$ and $6.97\kev$ line luminosities with respect to the continuum luminosity ($3-10\kev$). In all cases, a clear positive relationship is seen between the line and continuum luminosities. For $6.7\kev$ and $6.97\kev$ features, the relationship appears to be the same for NLS1 (blue) and BLS1 (red) galaxies, in line with the idea that their properties are indistinguishable between NLS1s and BLS1s. For the $6.4\kev$ feature however, the line features appear systematically dimmer for NLS1 galaxies compared to BLS1s. This is particularly prevalent for continuum values of log(L/erg s$^{-1}$) $\sim 43-44$ however, it should also be pointed out that these objects have large uncertainties.

Altogether, it appears that while the properties of ionised iron lines (\fexxv\ and \fexxvi) appear very similar between the two classes, differences appear in the narrow, neutral \feka\ lines, which are weaker and harder to detect in NLS1 galaxies.

\subsection{Model (B) - Broad emission line}
\label{sect:bline}

In model (B), the narrow \fexxv\ and \fexxvi\ features are removed and replaced with a single broad component, with line energy and width free to vary. The model still includes the narrow emission line at $6.4\kev$, and therefore simultaneously models reflection from the distant, neutral torus (narrow component) and from the inner accretion disc (broad component). As with model (A), a clear fit improvement compared to a power law model is seen for most objects, suggesting the presence of some form of iron emission in most of the sample. Additionally, the fit improvement is seen in both NLS1 and BLS1 galaxies, which suggests that both NLS1s and BLS1s exhibit these emission features, as expected (e.g. \citealt{Reeves+2006,Nandra+2007}). 

\begin{figure*}
	\centering
	\subfloat[]{
		\includegraphics[width=55mm]{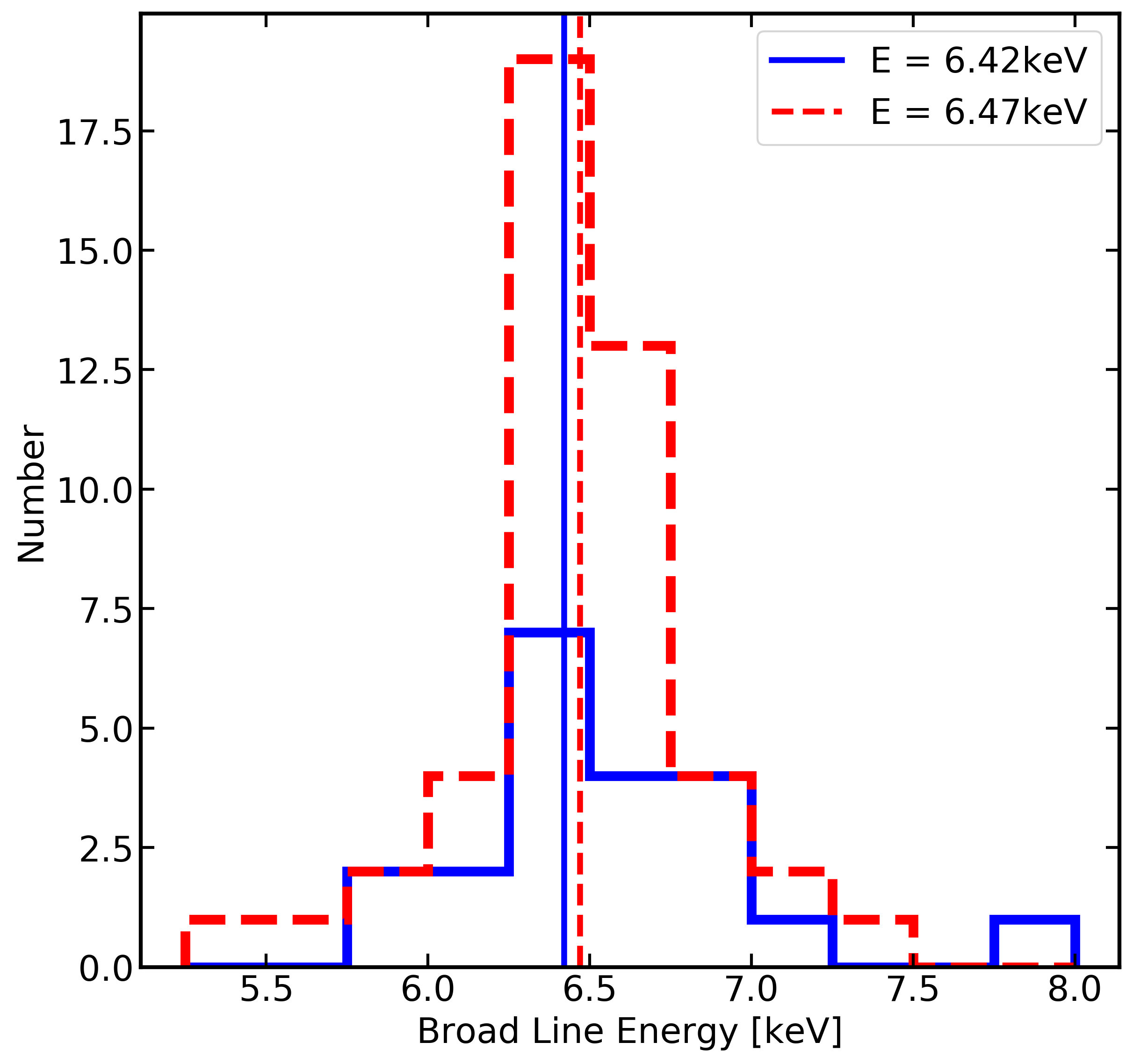}
	}
	\hspace{2mm}
	\subfloat[]{
		\includegraphics[width=53.5mm]{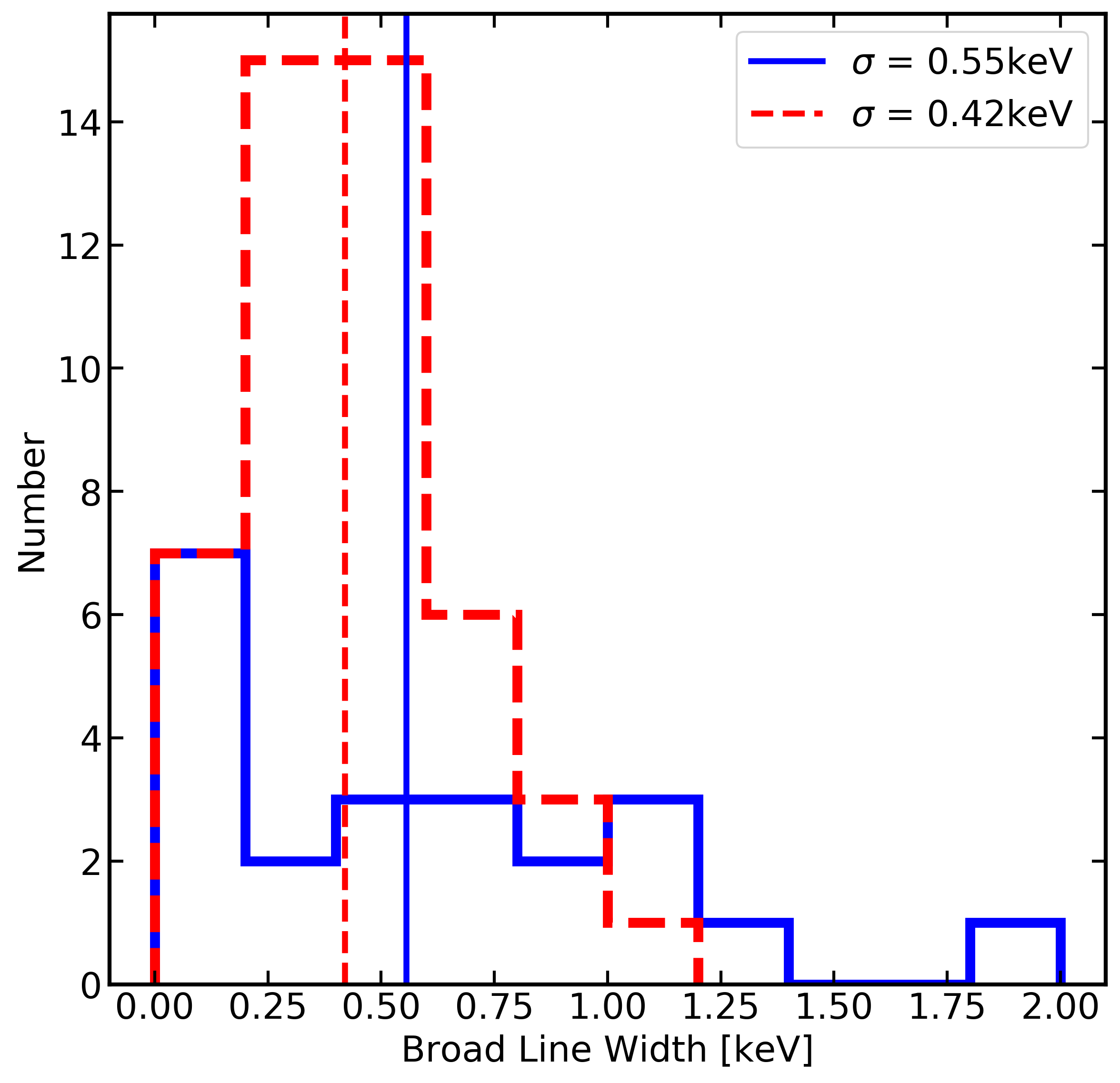}
	}
	\hspace{2mm}
	\subfloat[]{
		\includegraphics[width=55mm]{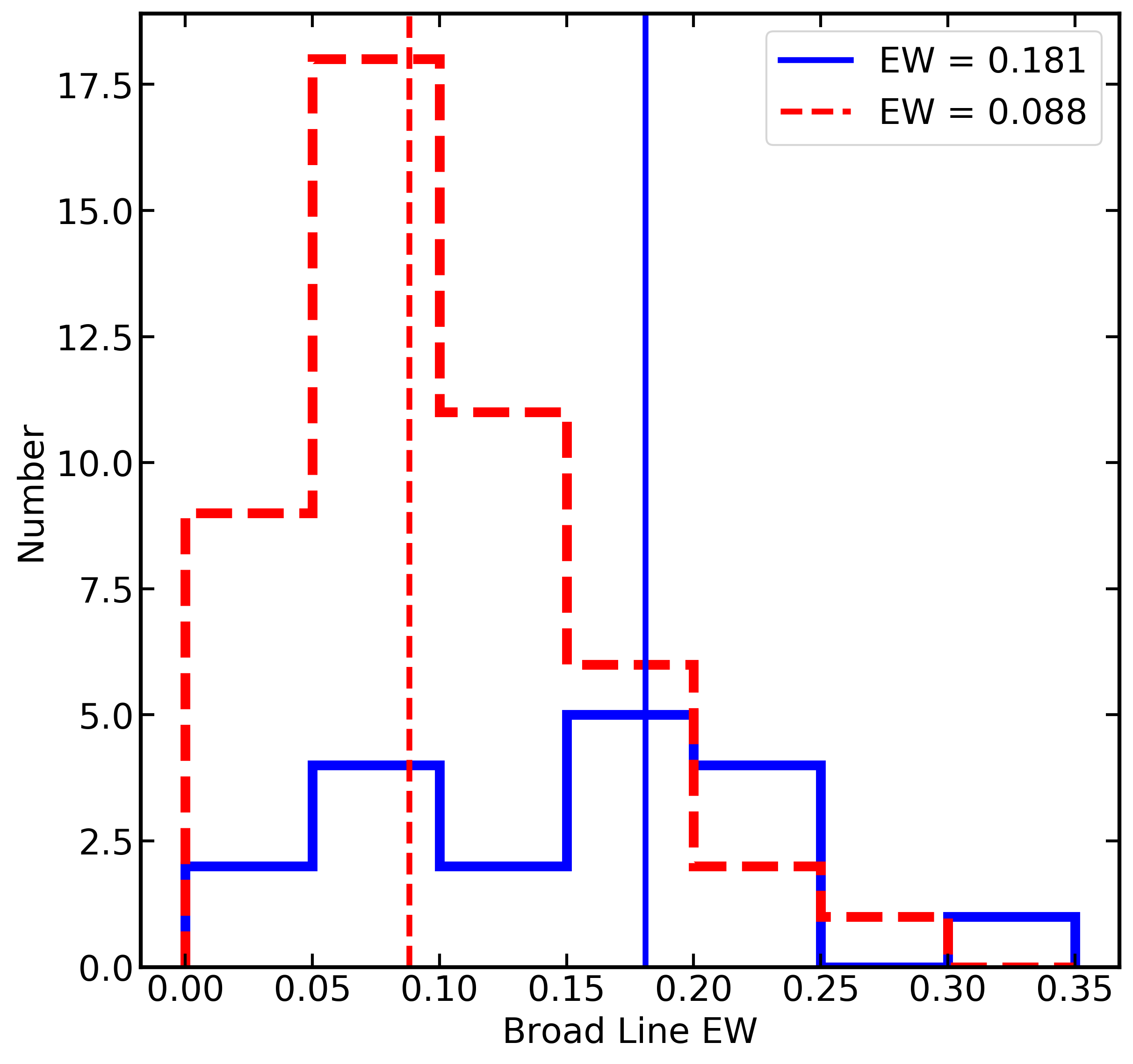}
	}
	\caption{Histograms showing parameter distributions, where the NLS1 parameter values are shown in blue solid lines and BLS1 parameter values are shown as red dashed lines. The median values for each parameter are shown as vertical lines in corresponding colours and line styles, and median values are given in the label for each plot. Panel (a) shows the broad line energy in units of keV (KS test p-value 0.50); (b) shows the broad line width in units of keV (p-value 0.079) and (c) shows the broad line equivalent width (p-value 0.00043). }
	\label{fig:blhist}
\end{figure*}

As before, it is useful to compare the detection rate for each feature between the two samples. Examiming first the broad component, 41/47 BLS1s (87\%) and 20/22 NLS1s (91\%) have an additional line with non-zero luminosity (at the 90\% confidence level). Additionally, 34/47 BLS1s (72\%) and 17/22 NLS1s (77\%) have widths constrained to be $>0$. Interestingly, these fractions are higher than those observed in some other works (e.g. \citealt{Reeves+2006,Nandra+2007}). The detection frequency between samples is very similar; this is sensible, as the high energy component of the broad line may model the same emission as the ionised features in Section~\ref{sect:nline} which were detected in a similar fraction of NLS1 and BLS1s. The lower energy tail of the broad line likely models the excess emission seen below $6.4\kev$ in some sources, such as Mrk 335 (see Fig.~\ref{fig:m335fe}). 

Comparing instead the narrow, $6.4\kev$ emission, 34/47 (72\%) of BLS1s show evidence for features with non-zero luminosity, while the same is true for only 8/22 (36\%) of NLS1s. For both samples, the number of objects for which the narrow $6.4\kev$ line luminosity is consitent with 0 is increased, likely due to the difficulty in separating the overlapping broad and narrow components. However, it is clear that the narrow core remains far weaker and more difficult to detect in NLS1s, independant of the model.

Next, histograms for some properties are compared in Fig.~\ref{fig:blhist}. NLS1s are shown in blue solid lines, and BLS1s are shown as red dashed lines. The median values for each parameter are shown as vertical lines in corresponding colours and line styles, and median values are given in the label for each plot. Panels (a) and (b) show the broad line energy and width, respectively. The median energies of $\sim6.4-6.5\kev$ and widths of $\sim0.5\kev$ suggest that  the broad components may be interpreted as iron emission which has been broadened in the accretion disc. The KS test p-values of 0.50 and 0.079 for the line energy and width, respectively suggest minimal differences in these parameters between the two samples. 

The NLS1 sample has a slightly lower median broad line luminosity, and performing a KS test suggests that the samples are drawn from slightly different distributions, with a p-value of 0.041. However, the NLS1s also have steeper photon indices, so the equivalent width should be used to provide a clearer view of any potential differences. Examining this instead (shown in panel (c) of Fig.~\ref{fig:blhist}) shows that the NLS1 sample has a much higher median equivalent width, with a very low p-value of 0.00043. This implies that the samples are different at the $>99.95\%$ confidence level, thereby suggesting that NLS1s have much stronger broad lines than their BLS1 counterparts. 

Examining the narrow line shows the opposite relationship between the samples. The median values for both luminosity and equivalent width are found to be lower for NLS1s, with the KS test yielding a p-value of 0.0027 and 0.0033, respectively. Compared to model (A), model (B) suggests even more significant distinctions between the narrow $6.4\kev$ emission line properties. 

\begin{figure*}
	\centering
	\subfloat[]{
		\includegraphics[width=0.95\columnwidth]{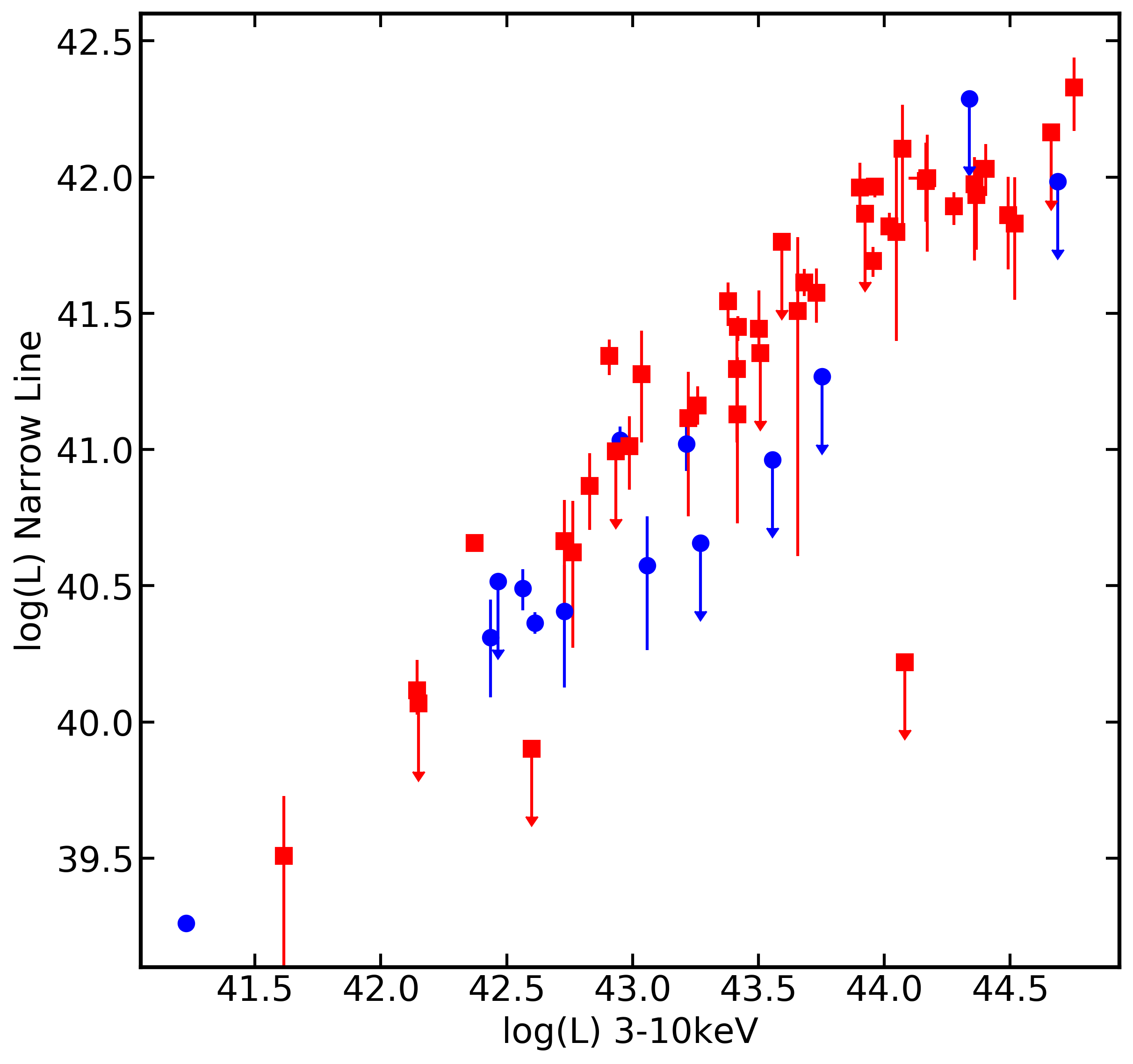}
	}
	\hspace{5mm}
	\subfloat[]{
		\includegraphics[width=0.95\columnwidth]{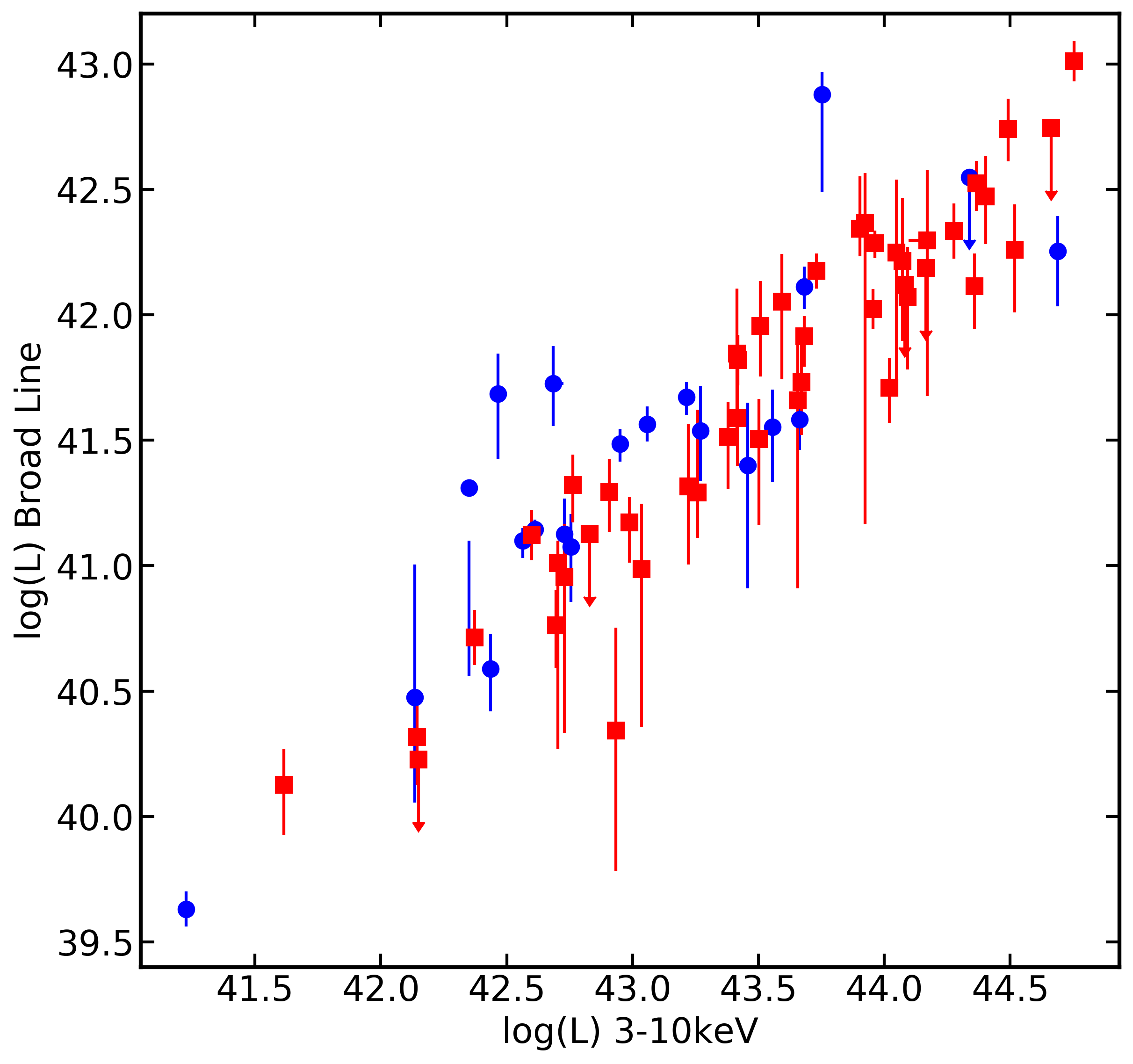}
	}
	\caption{Line luminosities (\ergps) with respect to the continuum luminosity for each source in the sample. Both values are shown as log values for clarity. NLS1 sources are shown as blue circles, and BLS1 sources are shown as red squares. Upper limits are indicated with arrows. NLS1 galaxies appear to show weaker narrow lines (panel a) but stronger broad lines (panel b) than BLS1s. }
	\label{fig:blsct}
\end{figure*}

As with model (A), it is also useful to compare the distributions of the line properties with those of the continuum. Fig.~\ref{fig:blsct} shows the distributions of the broad (a), and narrow (b) line luminosities with respect to the continuum luminosity. The distribution in panel (a) of Fig.~\ref{fig:blsct} is similar to that obtained with model (A), shown in panel (a) of Fig.~\ref{fig:nlsct}. As before, the luminosity of the narrow line appears shifted down compared to the BLS1 sample, with the most prominent differences between log(L/erg s$^{-1}$) $\sim 43-44$. This further highlights the idea that NLS1s have weaker narrow $6.4\kev$ lines, and implies that this trend is largely model independent. This, as well as the differences observed in properties of the broad line, will be explored further in Section~\ref{sect:discemis}.

Examining panel (b), there is some indication that NLS1s have brighter broad lines with respect to their continuua when compared to BLS1s, although the distinctions are less apparent than those seen by examining the equivalent width. Nevertheless, it appears that NLS1 galaxies have stronger broad lines than BLS1 galaxies.

\subsection{Model (C) - Absorption edge}
\label{sect:edge}

The absorption of a high-energy photon that ionizes the K-shell in neutral iron can result in the emission of a photon as a higher bound electron drops to fill the hole (fluorescence) or additional free elections through the Auger effect. The probability that a fluorescence photon is created depends on the fluorescence yield (e.g. \citealt{Bambynek+1972}), which is $\sim35$ per cent for neutral iron.

This process is examined with model (C), which includes a narrow feature at $\sim6.4\kev$ as well as an edge at $\sim7.1\kev$. Since the K-shell absorbs photons with energies higher than $7.1\kev$, the PIN data are included in order to provide spectral coverage from $\sim15-40\kev$. This allows for a more accurate measurement of the luminosity absorbed in the edge. As discussed in Section~\ref{sect:model}, the PIN data reveal signatures of a Compton hump for many sources. Following WG20, a broad Gaussian with an energy of $20\kev$ and a width of $5\kev$ is also added to the model for each source.

\begin{figure}
	\centering
	\includegraphics[width=0.95\columnwidth]{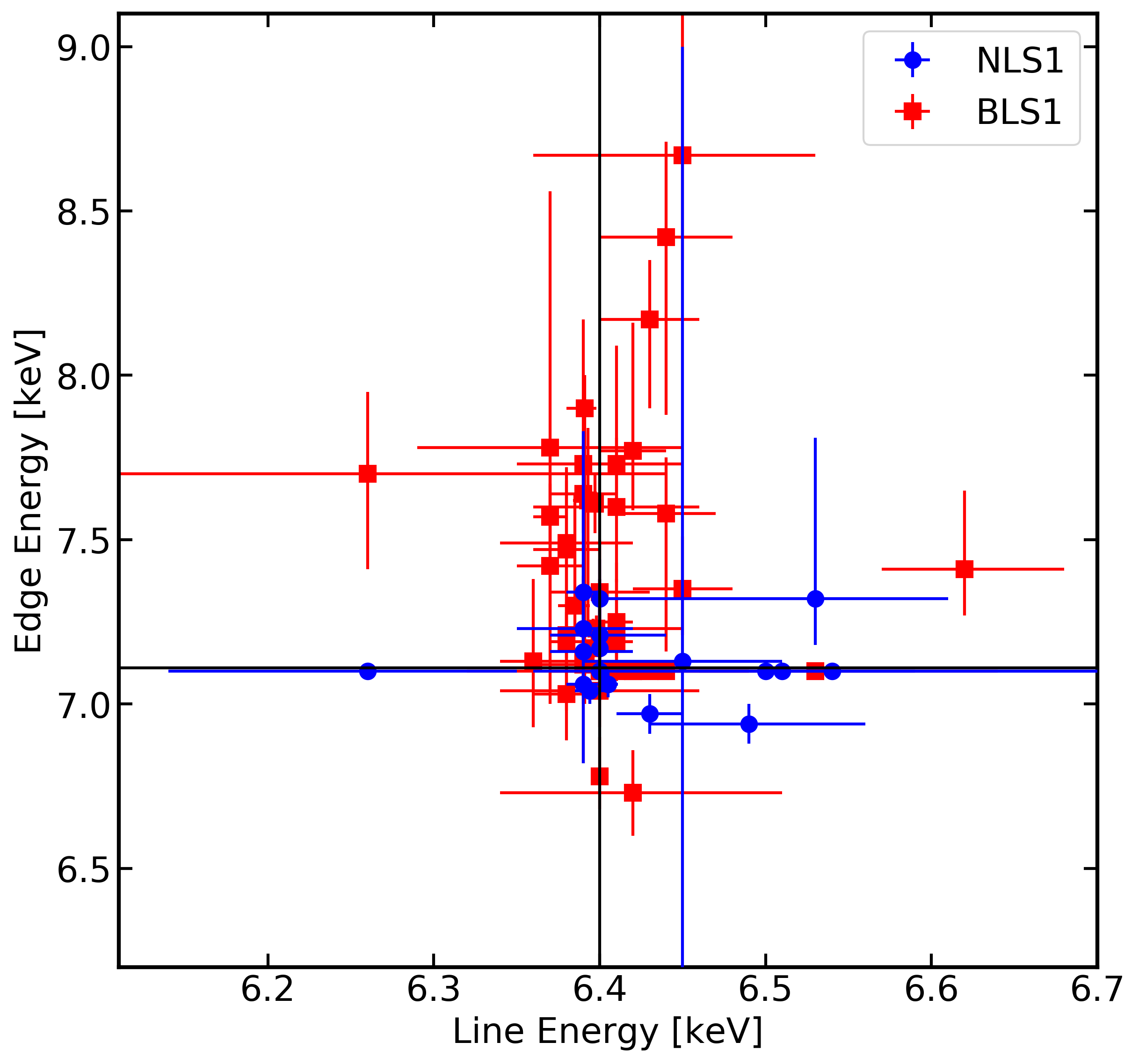}
	\caption{Best fit emission line and absorption edge energies for NLS1s (blue circles) and BLS1s (red squares). The expected energies of $6.4$ and $7.1\kev$ for the line and edge, respectively, are shown as black solid lines. }
	\label{fig:edge}
\end{figure}

To begin spectral fitting, the energies of the emission and absorption features are frozen at the expected values for neutral iron. The line and edge energies are then allowed to vary, and the C-statistic is re-evaluated to check for an improvement. This was the case for 60/69 sources, in which the average fit improvement was $\Delta$C $\sim8$ for 2 additional free parameters. The measured values are shown in panel (a) of Fig.~\ref{fig:edge}. The expected values of $6.4\kev$ for the emission line and $7.1\kev$ for the absorption edge are shown as solid black lines. NLS1 galaxies are shown as blue circles, while BLS1s are shown as red squares.

For most sources (48/69) the best fit line energies are in agreement with $6.4\kev$. Most NLS1s (15/22) have absorption edge energies that agree with $7.1\kev$.  However, many BLS1 galaxies (25/47) have edge energies that are significantly higher. The median edge energy is $7.16\kev$ for NLS1s, and $7.41\kev$ for BLS1s; and from Fig.~\ref{fig:edge}, it can be seen that some sources deviate by more than $1\sigma$ from the neutral energy value. Performing the KS test gives a p-value of 0.0056, suggesting that the edges in BLS1 galaxies are at significantly higher energies. Removing all the sources where the best fit edge energy is frozen, the p-value decreases to 0.0041, further suggesting potential differences. The optical depth ($\tau$) is comparable in both samples with $\tau\approx 0.1$.

As with the previous models, it is also of interest to compare the fraction of sources which show some evidence for an absorption edge. Doing so, it can be concluded that 12/22 (55\%)  NLS1s and 35/47 (74\%) BLS1s have $\tau>0$ at the 90 per cent level. While these fractions differ slightly between classes, this may be due to higher background in NLS1 spectra appearing because of the steeper spectra rather than from intrinsic difference between the classes.

Next, the luminosity of the emission line and the amount of radiation absorbed in the edge were measured. To estimate the absorbed luminosity, the difference in the $7.1-20\kev$ flux between the best-fit model and that with $\tau$ set to zero is measured. Errors on the luminosity absorbed by the edge are estimated using the measurement range on the optical depth.  

As with models (A) and (B), NLS1s once again show evidence for lower line luminosities, with a p-value of 0.0040. The amount of luminosity absorbed by the edge component is comparable between the classes. The KS test gives a p-value of 0.064, with the BLS1 sources having slightly higher median; however, at only the 90\% confidence level, it is difficult to claim that the distributions are truly different.  

We will refer to the ratio of the luminosity emitted in the iron line to the luminosity absorbed by the edge as the ratio of iron emission-to-absorption. ($\omega$). It is expected that in a system of pure fluorescence, $35$ per-cent of the photons absorbed in the $7.1-20\kev$ band will be re-emitted in the $6.4\kev$ band, as this is the fluorescence yield for iron (see for example, \citealt{Reynolds+2009}). However, the effects of different geometries, column densities and scattering may affect the measured values of the ratio of iron emission-to-absorption (e.g. \citealt{MY+2009}), such that the they may differ from the true value of $0.35$. This is discussed in detail in Section~\ref{sect:discedge}.

\begin{figure}
	\centering
		\subfloat[]{
		\includegraphics[width=0.95\columnwidth]{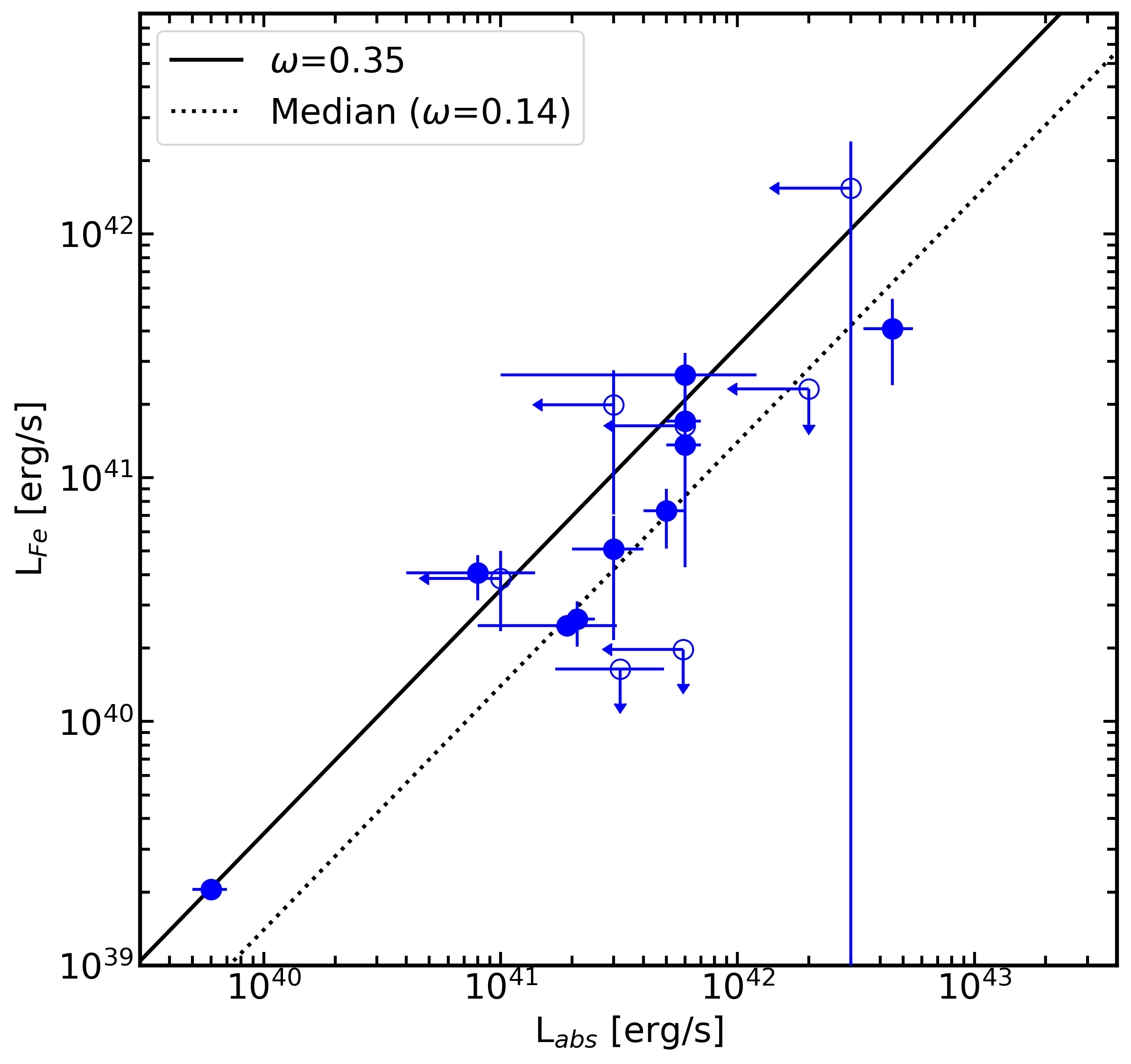}
	}
	\hspace{5mm}
	\subfloat[]{
		\includegraphics[width=0.95\columnwidth]{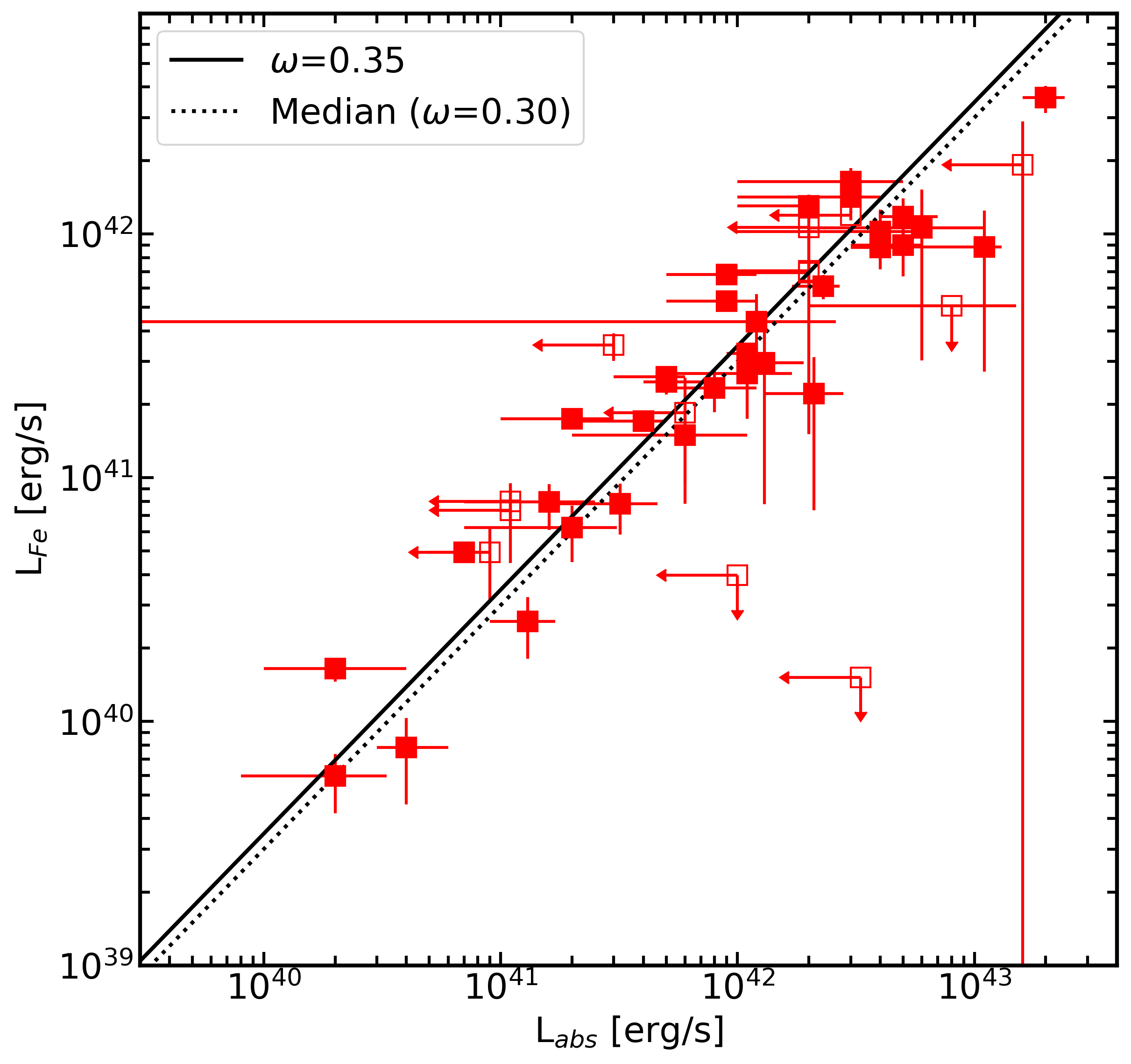}
	}
	\caption{The distributions of iron line luminsosity (L$_{\rm Fe}$) and luminosity absorbed by the edge component (L$_{\rm abs}$). NLS1s are shown in panel (a) in blue, and BLS1s are shown in panel (b) in red. Upper limits are indicated both with arrows, and with unfilled data points, to clear show which sources have only upper limits measured for one of both parameters. A fluorescence yield of $0.35$ is shown as a solid black line, and median ratios of iron emission-to-absorption for each sample are shown as black dotted lines.} 
	\label{fig:fy}
\end{figure}

Fig.~\ref{fig:fy} shows the values for the absorbed luminosity (L$_{\rm abs}$) and the iron line luminosity (L$_{\rm Fe}$). NLS1 galaxies are shown as blue circles in panel (a), while BLS1s are shown as red squares in panel (b). Upper limits are indicated both with arrows, and with unfilled data points, to clearly show which sources have only upper limits measured for one of both parameters. The fluorescence yield of $0.35$ is shown as a solid black line, and the median ratio of iron emission-to-absortion for all filled in points are shown as black dotted lines. While most sources seem to agree with an $\omega$ of $0.35$, at least within error, some deviate significantly from the expected value. It appears that some NLS1s galaxies have weaker iron lines than is expected for the amount of continuum luminosity absorbed (e.g. lie below the line). For the NLS1 sample, the average ratio is L$_{\rm Fe}$ /  L$_{\rm abs} \sim 0.14$. For BLS1s, the same ratio is $\sim0.30$, in agreement with what is expected from the theoretical fluorescence yield.

Performing the KS test gives a p-value of 0.0064, implying that the distributions are significantly different. When removing the sources for which the iron emission and absorption could not be constrained, both median values increase slightly, to 0.36 for BLS1s and 0.20 for NLS1s. The p-value also increases to 0.12, thus slightly decreasing the confidence that the two samples are drawn from very different distributions. Nevertheless, the marked differences between the two samples, and the fact that the median measured ratio of iron emission-to-absorption for NLS1s is well below the expected fluorescence yield value of 0.35, is worthy of further investigation.

\section{Discussion}
\label{sect:disc}

\subsection{The Fe~K band in the \suzaku\ sample of Seyfert 1s}
Using the models presented in this work, different iron emission properties can be measured, and the results for NLS1 and BLS1 galaxies compared. To truly model the emission from the AGN, a model should include a narrow \feka\ component, ionised emission lines, a broad component, and an absorption edge, as all of these components arise from different physical origins and are likely observed simultaneously. However, when combining the components of two or three models used in this work and applying these to even the brightest sources in this sample, it becomes difficult or impossible to constrain any of the iron line fluxes or edge parameters. In particular, disrepancies exist when modelling the absorption edge and broad component, as both manifest in a sharp spectral dip at $\sim7\kev$ present in many sources (e.g. \citealt{Boller+2002,Boller+2003,Nandra+2007,Miller+2008,Reynolds+2009}), and distinguishing between such features would likely require deep observations with calorimeters, such as with \xrism\ (\citealt{xrismins}). By contrast, fitting each model individually allows for good constraints on many relevant parameters, and comparison between sample properties within each model and between models is possible.

Furthermore, some of the differences observed in this work between line luminosities in NLS1 and BLS1 galaxies may be related to luminosity discrepancies between the two samples. This may explain why the differences in average properties can be more pronounced when considering line luminosities compared to equivalent widths. In WG20, the median luminosities between NLS1 and BLS1 galaxies are compared, and the use of the KS test shows no evidence that the luminosity distributions differ between samples (p-value 0.12; e.g. see Fig. 5, panel (a) and section 4.4 of WG20), although the median X-ray luminosity is slightly higher (factor $\sim3$) for BLS1s. Therefore, it seems unlikely that luminosity differences have a strong effect on the observed distributions, but this cannot entirely be ruled out.

Nevertheless, the analysis presented in this work revealed numerous interesting similarities and differences in the iron emission and absorption properties of NLS1 and BLS1 galaxies. The main results can be summarised as follows:

\begin{enumerate}[(i)]
	\item Model (A) shows that NLS1s and BLS1s have very similar ionised lines. NLS1s have slightly more luminous $6.7\kev$ lines, and BLS1s have slightly more luminous $6.97\kev$ lines, but these differences are not statistically significant. This suggests a common origin for these features, such as the BLR or ionised gas in the inner layers of the torus.
	
	\item In model (B), NLS1s have significantly larger broad line equivalent widths than BLS1s. This difference is significant at the $>99.95\%$ ($\sim3.5\sigma$) confidence level, and thus implies a significant difference between NLS1 and BLS1 spectra.
	
	
	\item Independent of the model tested (i.e. A, B and C), the narrow neutral \feka\ emission feature is weaker in NLS1s than BLS1s. This is true when comparing detection rates, line luminosities and equivalent widths. This suggests that weaker neutral narrow lines is an intrinsic charasteric for NLS1s and not merely a result of modeling errors or biases. 
	
	\item NLS1 and BLS1 galaxies are better fit with model (C) when the line and edge energy are left free to vary. While the line energies typically agree with $6.4\kev$ for both samples, BLS1 galaxies have significantly higher edge energies than NLS1s. The median value is $7.41\kev$, which may indicate intermediate ionisation states of the iron atoms for these sources.
	
	\item The measured raio of iron emission-to-absorption is lower in NLS1s than in BLS1s. The BLS1 median ($\sim0.3$) agrees with the theoretical value of the fluorescence yield ($\sim0.35$), however, the median is much lower for NLS1s ($\sim0.14$).
	
\end{enumerate}

\begin{figure}
	\centering
	\includegraphics[width=0.95\columnwidth]{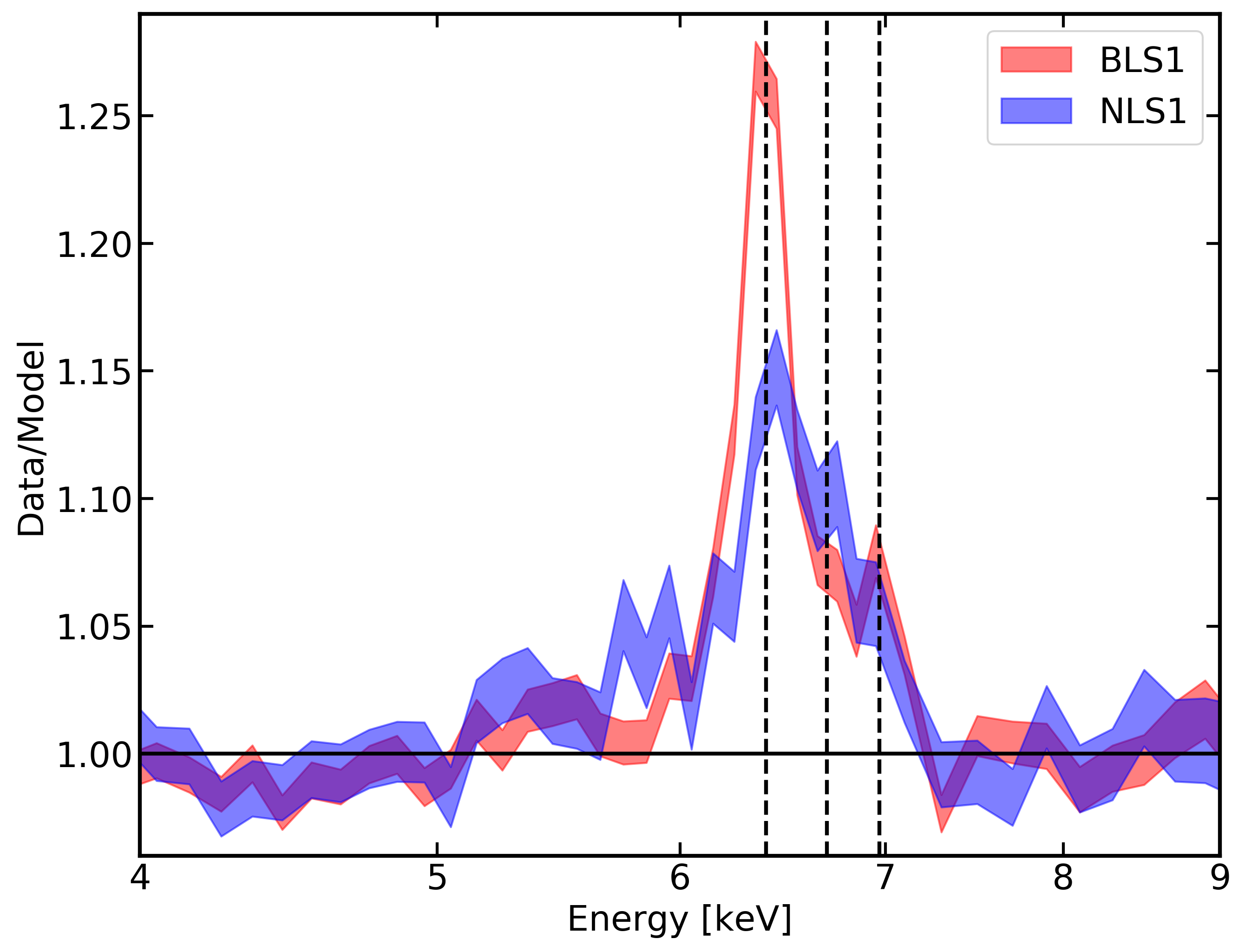}
	\caption{Stacked residuals from a power law fit to all BLS1 and NLS1 spectra. The stacked BLS1 data are shown in red, and NLS1s are shown in blue. Black dashed lines show energies of $6.4$, $6.7$, and $6.97\kev$, where iron features are expected. }
	\label{fig:stack}
\end{figure}

Most of the properties above are illustrated in Fig.~\ref{fig:stack}, where the stacked residuals from fitting a power law to each individual spectrum (e.g. \citealt{Chaudary+2012}) are presented. NLS1s and BLS1s were averaged individually, with the resulting stacked residuals spectrum for BLS1s shown in red, and for NLS1s shown in blue. Black dashed lines show energies of $6.4$, $6.7$, and $6.97\kev$, where iron features are expected. Examining this, the stacked NLS1 residuals show a weaker narrow \feka\ component, slightly stronger $6.7\kev$ emission, weaker $6.97\kev$ emission, and increased residual flux around $\sim5.5-6\kev$, which likely results from the higher median equivalent width of the broad line discussed in the previous sections.

Additionally, the stacked BLS1 residuals show some evidence for a dip at the expected energy of about $7.3-7.4\kev$. This is likely the edge component seen in many spectra, as the energy agrees with the median BLS1 edge energy from model (C). Curiously, the edge feature is much less well defined in the NLS1 spectra. This is likely due to the fact that there are fewer NLS1s being stacked compared to BLS1s rather than an intrinsic property.

\subsection{The presence of broad emission lines}
\label{sect:discemis}

Significant differences in the equivalent width of the broad line incorporated in model (B) exist between NLS1 and BLS1 AGN. To assess the robustness of these measured differences, the equivalent width is checked against other model parameters to search for correlations. The equivalent width does not appear to be correlated with line energy or with photon index ($\Gamma$). This suggests the stronger equivalent widths are not simply an artefact of the steeper continuum spectra typical of NLS1s. Rather, the larger broad line equivalent widths observed in NLS1s appear to be an intrinsic property of these sources.


Given that the equivalent width of the broad line component is significantly higher for the NLS1 sample, it is apparent that broad features and spectral complexity in the $4-7\kev$ region are a prominent element in NLS1 spectra, and it is sensible that these sources may more often better fit with a model including a broad line (e.g. \citealt{Gallo2006,Gallo+2018}).

The high equivalent width observed in NLS1s can give some indication as to the physical model that is best for these sources. These broad lines are typically associated with a blurred reflection model, in which a $6.4\kev$ feature is broadened due to general relativistic effects in the inner accretion disc (\citealt{Fabian+2004,RossFabian+2005}). The broad line energies of $6-7\kev$ and widths of $\sim0.5\kev$ seem to agree with this interpretation. It is also consistent with the stronger soft excesses and Compton humps found in NLS1s (WG20). It is therefore plausible that a blurred reflection component may have more significant contributions in NLS1 spectra than in BLS1 spectra. \cite{Zhou+2010} study the results from the sample presented in \cite{Nandra+2007}, and also find higher broad line EW in NLS1s; this is interpreted as blurred reflection, and in particular, smaller inner disc radii in NLS1s. This interpretation seems to be  in good agreement with the results from this work.

An alternative interpretation for the broad component may be curvature produced by absorption in a partial covering model (e.g. \citealt{Tanaka+2004}). Under this interpretation, optically thick material close to the corona absorbs some of the primary emission, producing curvature around $4-7\kev$ and a deep edge at $\sim7.1\kev$. As seen in Section~\ref{sect:edge}, many NLS1s show strong absorption edges between $7-7.5\kev$, so a partial covering model with higher absorber column density or covering fraction may explain the higher equivalent widths seen in NLS1s. This interpretation however may be at odds with the significant positive correlation seen between the hard and soft excess strengths presented in WG20, as simulations have shown that while blurred reflection produces a positive correlation, partial covering produces an anti-correlation, or no correlation between these parameters (\citealt{Vasudevan+2014}).

\subsection{Iron emission and absorption properties}
\label{sect:discedge}

The investigation performed using model (C) in Section~\ref{sect:edge} provides an opportunity to speculate on the nature of the absorbing material. The median edge energy is slightly higher in BLS1s ($\sim7.41\kev$) than in NLS1s ($\sim7.16\kev$). This might point to different ionisation states of the iron in the two samples, but the CCD resolution makes it difficult to determine this with significance. Regardless, the values of both are consistent with near neutral material and the expected fluorescence yield will be approximately $0.35$ in both samples. The potential differences in edge energies will be best-studied with large samples of AGN observed with calorimeter-type resolution (e.g. \citealt{xrismins}).

To investigate this further, the results from \cite{MY+2009} are considered, which present a relationship between observed \feka\ equivalent width and other physical parameters of the system, assuming that the emission and absorption originate from pure fluorescence. In particular, it depends on the solid angle of the torus as seen by the ionising source, the fluorescence yield, the iron abundance, the Fe~K absorption cross-section, the photon index of the ionising continuum ($\Gamma$), and the column density (see equation 5 of \citealt{MY+2009}). 

To get a sense of which of these properties, if any, differ between NLS1s and BLS1s, a ratio can be taken by considering the average properties for NLS1s and BLS1s (Equation 1). The absorption cross section is not expected to be different in NLS1s and BLS1s, so the ratio of this parameter will be one. The equation is further reduced with two assumptions. Firstly, while there is some evidence that some NLS1 galaxies have over-abundances of iron in the inner disc (e.g. \citealt{Jiang+2019}), abundance differences in the torus have not been reported so this ratio is 1.  Secondly, from our measurement in Section~\ref{sect:edge}, the optical depth is comparable in NLS1s and BLS1s.  Since $\tau$ is proportional to the column density, it appears that this is also comparable in both samples. Substituting these values into equation (5) of \cite{MY+2009} and rearranging, the following is obtained: 

\begin{equation}
\frac{EW_N}{EW_B} = \left( \frac{\Delta \Omega_N}{\Delta \Omega_B} \right) \left( \frac{\omega_N}{\omega_B} \right) \left( \frac{\Gamma_B + 1.67}{\Gamma_N + 1.67} \right)  \left(0.8985 \right)^{\Gamma_N - \Gamma_B}
\end{equation}

\noindent where $EW$ is the equivalent width, $\Delta \Omega$ is the solid angle (which is proportional to the opening angle of the torus), $\omega$ is the fluorescence yield of iron, $\Gamma$ is the photon index and the subscripts $N$ and $B$ indicate NLS1s and BLS1s, respectively.

The photon index ($\Gamma$) and the equivalent width of the iron line for each source has been measured, so ratios of the average values can be determined. For the equivalent width, this ratio is $\sim0.68$, and for all terms including $\Gamma$, the value is $\sim0.92$. The theoretical value for the fluorescence yield is governed by physics, and should be $\sim0.35$ for both NLS1s and BLS1s. Therefore, the ratio $\frac{\omega_N}{\omega_B}$ should be unity. This reduces Equation 1 to:

\begin{equation}
\left( \frac{\Delta \Omega_N}{\Delta \Omega_B} \right) \approx 0.74 
\end{equation}

Given that $\Delta \Omega$ is proportional to the cosine of the opening angle (or height) of the torus, this would imply that the opening angle of the torus for NLS1s is $\sim25\%$ larger than for BLS1s. This is a clear indication of potential differences in the torus, and can be interpreted as the NLS1 torus being closer to a disc shape, while the torus in BLS1s is closer to spherical. A possible physical mechanism to produce this difference is related to the accretion rate of the two samples. NLS1s accrete at a higher rate on average, and may therefore generate stronger winds that can evacuate more gas and dust from the central region, pushing it outwards, away from the AGN (e.g. \citealt{Takeuchi+2013}). This would result in more material from the torus being pushed away, causing the opening angle to increase, satisfying the equation and explaining the observed differences in iron properties.

The differences in measured median ratios of iron emission-to-absorption ($\sim0.14$ for NLS1s and $\sim0.30$ for BLS1s) may also be explained emission and absorption are arising in different mediums. Some of the \feka\ line emission in NLS1s may be produced in the BLR, rather than in the torus. Since an \feka\ line originating in the BLR may not necessarily produced via fluorescence, but rather through collisional effects or recombination, the emission-to-absorption ratio will differ. We also noted that NLS1s have much higher equivalent widths of the broad line. It is possible that NLS1s have much stronger spectral contributions from blurred reflection (e.g. Section~\ref{sect:discemis}). As the broad component is not present in model (C), it is possible that the edge arises from absorption in the torus and the disc (e.g. \citealt{RossFabian+2005}), but the narrow \feka\ emission only accounts for emission from the torus.  In this case, the measured $\omega$ would be underestimated in NLS1s. Additionally, absorption in the upper layers of the accretion disc may be present, deepening the observed edge or introducing variability (e.g. \citealt{GalloFabian+2011,GalloFabian+2013,Fabian+2020}).

It is not possible in this work to distinguish if the differences in equivalent width and the iron emission-to-absorption ratio arises from geometry, or from effects beyond pure fluorescence. Ultimately, it is likely to be a combination of these effects which produces the observed properties. One must also appreciate the large uncertainties in these values, and source-to-source variations. However, it is intriguing that the differences exist that might point to different torus geometries and/or increased contribution from the accretion disc and BLR to the iron emission in NLS1s. These scenarios can likely be understood in systems that are accreting at higher fractions of their Eddington limits.

\section{Conclusions}
\label{sect:conclusion}

This work presents an expansion of a previous study, WG20, by analysing the iron line profiles of 22 narrow-line Seyfert 1 and 47 broad-line Seyfert 1 galaxies observed with \suzaku. Three different models were used to characterise the iron emission lines and absorption edge in each source. It is likely that all three models should be used simultaneously to truly characterise the spectral features in AGN. However, the three models applied here allow for the proper characterisation of multiple components.

The analysis revealed numerous differences in iron emission and absorption properties between NLS1 and BLS1 samples. Weaker narrow \feka\ lines are observed in NLS1s in all three models. Stronger broad lines (higher equivalent widths) are also found in NLS1s , which may support a blurred reflection interpretation, in agreement with the results of WG20. Higher quality data with current and future missions including \xmm\ (\citealt{Jansen+2001}), \xrism\ (\citealt{xrismins}) and \athena\ (\citealt{athenains}) may help to confirm these trends, and spectral modelling with physical models such as blurred reflection and partial covering may shed further light on differences in the broad line properties.

Iron emission and absorption modelling also suggests differences in the equivalent width or the narrow neutral iron line, and in the ratio of iron emission-to-absorption. This may suggest a variety of physical properties, including differences in the geometry (in particular, the opening angle) of the torus, or emission aside from pure fluorescence. Ultimately, the differences could be attributed to a combination of these effects or different process in individual objects. Future broad-band X-ray spectral analysis with \xrism\ and \nustar\ (\citealt{Harrison+2013}) may help to probe the true physical origins of the observed distributions.

\section*{Acknowledgements}
This research has made use of data obtained from the \suzaku\ satellite, a collaborative mission between the space agencies of Japan (JAXA) and the USA (NASA). SGHW and LCG acknowledge the support of the Natural Sciences and Engineering Research Council of Canada (NSERC). LCG acknowledges financial support from the Canadian Space Agency (CSA).

\section*{Data Availability}
The data used in this work are publicly available in the \suzaku\ DARTS archive (https://darts.isas.jaxa.jp/astro/suzaku/data/).

%

\bibliographystyle{mnras}
\bibliography{dec21_iron_final}



%



\bsp	
\label{lastpage}
\end{document}